\pdfoutput=1
\documentclass[aps,prl,twocolumn,superscriptaddress]{revtex4-1}
\usepackage[colorlinks=true,citecolor=blue,urlcolor=blue,linkcolor=blue,bookmarksopen]{hyperref}
\usepackage{amsmath,amssymb}
\usepackage{bm}
\usepackage{graphicx}
\usepackage{comment}

\renewcommand\d{\partial}
\newcommand\<{\langle}
\renewcommand\>{\rangle}
\newcommand\x{{\mathbf{x}}}
\newcommand\p{{\mathbf{p}}}
\newcommand\q{{\mathbf{q}}}
\renewcommand\j{{\mathbf{j}}}
\newcommand\J{{\mathbf{J}}}
\newcommand\E{{\mathbf{E}}}
\newcommand\A{{\mathbf{A}}}
\newcommand\B{{\mathbf{B}}}
\newcommand\z{\mathbf{\hat z}}
\renewcommand\v{{\bm{v}}}

\renewcommand\Re{\mathop{\mathrm{Re}}}
\renewcommand\Im{\mathop{\mathrm{Im}}}
\newcommand\+{\dagger}
\newcommand\waa{w_{\rm AA}}
\newcommand\wab{w_{\rm AB}}
\newcommand{\vp} {\varphi}
\def\cS{\mathcal{S}}
\newcommand{\Laplace}{\bigtriangleup}

\newcommand{\ve} {\varepsilon}

\begin{document}
\title{Electrodynamics of Thin Sheets of Twisted Material}
\author{Dung Xuan Nguyen}
\affiliation{Brown Theoretical Physics Center and Department of Physics, Brown University, 182 Hope Street, Providence, Rhode Island 02912, USA}
\author{Dam Thanh Son}
\affiliation{Kadanoff Center for Theoretical Physics, University of Chicago,
  933 East 56th Street, Illinois 60637, USA}

\begin{abstract}

We construct a minimal theory describing the optical activity of a
thin sheet of a twisted material, the simplest example of which is
twisted bilayer graphene.  We introduce the notion of ``twisted
electrical conductivity,'' which parametrizes the parity-odd response
of a thin film to a perpendicularly falling electromagnetic waves with
wavelength larger than the thickness of the sheet.
We show that the
low-frequency Faraday rotation angle has different
behaviors in different phases.  For an insulator, the Faraday angle
behaves as $\omega^2$ at low frequencies, with the coefficient being
determined by the linear relationship between a component of the
electric quadrupole moment and the external electric field.  For
superconductors, the Faraday rotation angle is constant when the
frequency of the incoming EM waves is below the superconducting gap
and is determined by the coefficient of the Lifshitz invariant in the
Ginzburg-Landau functional describing the superconducting state.  In
the metallic state, we show that the twisted conductivity is
proportional to the ``magnetic helicity'' (scalar product of the
velocity and the magnetic moment) of the quasiparticle, averaged around
the Fermi surface.  The theory is general and is applicable to
strongly correlated phases.

\end{abstract}
\maketitle

\emph{Introduction.}---Recently, twisted quasi-two-di\-men\-sion\-al
materials have attracted intense intention.  The most well-known
example of such materials is twisted bilayer graphene, where flat
bands have been predicted and found and correlated insulating and
superconducting states observed
\cite{Bistritzer2011,LopesdosSantos2007,SurezMorell2010,LopesdosSantos2012,Cao2018,Cao2018-insulator}.
Flat bands have also been discovered in the other twisted systems
\cite{Carr2018,Kariyado2019}, including twisted double bilayer
graphene~\cite{Shen2020,Liu2020,Cao2020} and twisted trilayer
graphene~\cite{Tsai2019-trilayer}.

One probe of the states of the twisted quasi-2D materials is through
their interaction with the external electromagnetic field.  Due to the
twisted structure, a circularly polarized electromagnetic wave passing
through such a material can experience different responses depending
on the sign of the polarization---the so-called Faraday rotation and
circular dichroism~\cite{Kim2016,Addison2019}.  In this paper we study
the question: what does the chiral response of a twisted quasi-2D
material to long-wavelength electromagnetic waves reveal about the
structure of the low-energy excitations in such a materials?

We first construct here a minimal theory that describes the chiral
response of a thin sheet of material with thickness $d$ much smaller
than the wavelength $\lambda$ of the incoming light.  We show that in
addition to the usual 2D electrical conductivity, one just needs a new
kinetic coefficient which we dub ``twisted conductivity,'' which
determines the electric current generated by a gradient of the
electric field, or the time dependence of the magnetic field.  The
same quantity also determines the ``dipole current,'' to be defined,
in an external electric field.

In the metallic phase this twisted conductivity is shown to be
proportional to the ``magnetic helicity'' of the quasiparticle
excitations on the Fermi surface, defined as the product of the
velocity and the in-plane magnetic moment of the Landau quasiparticle.

Previous theoretical work include
Refs.~\cite{SurezMorell2017,Stauber2018,Ochoa2020}, where one
considers a model of two layers, the current on one layer is linearly
dependent on the electric field on that layer and on the other
layer. This approach is limited to two-layer systems within the
approximation that the electrons are localized in a very thin shell
around each layer, whose thickness is much less than the distance
between the layers.  In Ref.~\cite{Wang2020} the Kubo formula for the
chiral response is derived for commensurate twist angles.

\emph{General consideration.}---Consider a thin sheet of material,
stretched along the $(x,y)$ directions.  We will have in mind twisted
bilayer graphene as a prototype.  We are interested in the chiral
response of this materials to electromagnetic waves which have
wavelength much larger than the thickness of the sheet.  For example,
we want to understand Faraday's rotation of the plane of polarization
of an incoming electromagnetic wave.

We will be mostly interested in the case when the incoming
electromagnetic waves fall perpendicularly to the plane of the sheet.
In this case the is electric field parallel to the sheet.

We assume that the sheet is invariant under a discrete group of
rotations around the $z$ axis, sufficiently large for the conductivity
tensor to be isotropic (for example $C_{3z}$ in the case of twisted
graphene sheets).  The twisted sheet is assumed to be not invariant
under 2D reflection but only under a 2D reflection (say, $x\to x$,
$y\to-y$) combined with $z\to-z$, which is simply the two-fold
rotation around the $x$ axis $C_{2x}$.  This symmetry is present in
twisted bilayer graphene \cite{Balents2019}.  We will assume that this symmetry is not
spontaneously broken.  We will also assume time reversal invariance.

In the standard treatment, the response of a single thin layer (say, of
graphene) to an incident electromagnetic wave is characterized by the
2D (complex) conductivity
$\sigma(\omega)$~\cite{Nair2008,Mak2008,Li2018}.  To treat effects
like Faraday rotation or circular dichroism, the surface conductivity
is not sufficient.  We now develop a formalism, which deals
exclusively with two-dimensional quantities, but still allows one to
captures these effects.

Let us denote the current in a finite-width piece of material by
$\J(z,\x)$.  If the sheet is thin, we can define the 2D current as
\begin{equation}
  \j(\x) = \int\!dz\, \J(z,\x),
\end{equation}
and the linear electric response is given by the frequency-dependent
electrical conductivity: $\j=\sigma\E$.  The transmission and
reflection amplitudes of an electromagnetic wave falling
perpendicularly onto the sheet then can be expressed through the real
and imaginary parts of $\sigma$~\cite{Li2018}.
To go beyond $\sigma$, we introduce a new current
\begin{equation}
  \bm\zeta (\x) = \int\!dz\, z \J(z,\x).
\end{equation}
For a system which consists of two layers separated by a distance $d$,
if the thickness of the electron orbitals on each of the layers is
much smaller than $d$, then $\bm{\zeta}=\frac d2 (\j_1-\j_2)$, where
$\j_1$ and $\j_2$ are the electric currents on the layers $1$ and $2$.
We term this quantity the ``dipole current,'' because if there is no
tunneling between the two layers, then the electric dipole moment
along the $z$ direction is conserved, and $\bm{\zeta}$ is the current
of that conserved charge.

The source of the current $\j$ are the electric field $\E$, and the
source of the dipole current $\bm\zeta$ is the gradient of the electric
field along the perpendicular direction $\d_z\E$.  This can be seen by
noticing that the Ohmic heat generated in the system per unit area is
\begin{equation}
  \int\!dz\, \J(z)\cdot \E(z)
  \approx \!\int\!dz\, \J(z)\cdot (\E + z \d_z \E)
  =  \j \cdot \E + \bm{\zeta} \cdot \d_z \E .
\end{equation}
We now introduce the conductivities $\sigma$, $\sigma_1$,
$\tilde\sigma_1$, and $\sigma_2$:
\begin{align}
  \j &= \sigma \E - \sigma_1 \d_z \E \times \z , \\
  \bm\zeta &=  \tilde\sigma_1 (\E\times \z) + \sigma_2 \d_z\E .
\end{align}
In general the conductivities are functions of the frequency.  Time
reversal invariance implies the Onsager relation
$\tilde\sigma_1(\omega) = \sigma_1(\omega)$.  Positivity of entropy
production implies that both $\sigma$ and $\sigma_2$ have positive
real parts, and $\Re\sigma\Re\sigma_2 \ge (\Re\sigma_1)^2$.

On dimensional ground, we expect that typically $\sigma_1 \sim
d\sigma$, $\sigma_2 \sim d^2\sigma$.  For electromagnetic waves with
wavelength $\lambda$, the effects of $\sigma_1$ and $\sigma_2$ will be
suppressed by $d/\lambda$ and $(d/\lambda)^2$, respectively.  However,
since $\sigma_d$ is the first transport coefficient that breaks
reflection symmetry, we need to keep it in order to compute, e.g.,
Faraday rotation.  Although the effect $\sigma_2$ is always small, we
will however keep it in our formalism for the sake of symmetry.

Considering an electromagnetic plane wave with frequency $\omega$
falling perpendicularly onto the plane.  We can work in the gauge
where $A_0=A_z=0$, where the Maxwell equation for the perpendicular
components of the vector potential is (we use the Gauss
units)
\begin{equation}\label{Maxwell-3d}
  - \d_z^2 \mathbf{A} - \frac{\omega^2}{c^2}\mathbf{A} =
  \frac{4\pi}c \J.
\end{equation}
(We assume here that the thin sheet is immerse in vacuum.  Our calculation
can be trivially modified when the medium on one or both
sides of the sheet has a nontrivial dielectric constant.)  To find the
boundary conditions at $z=0$ we can integrate over $z$ from
$z=-\epsilon$ to $z=+\epsilon$
\begin{equation}\label{bc1}
  - \A'|^\epsilon_{-\epsilon} = \frac{4\pi}c \j = \frac{4\pi i\omega}{c^2}
    (\sigma \A - \sigma_1 \A'\times \z). 
\end{equation}
We can also multiply Eq.~(\ref{Maxwell-3d}) by $z$ and integrate over
$z$ to find
\begin{equation}\label{bc2}
  \delta \A = \frac{4\pi}c \bm{\zeta} =  \frac{4\pi i\omega}{c^2}
  (\sigma_1\A\times\z + \sigma_2 \A')  ,
\end{equation}
where we have denoted the jump of a quantity $A$ across the sheet by
$\delta A\equiv A(+\epsilon)-A(-\epsilon)$.  These boundary conditions
can be written for the two circular polarizations $A_\pm=A_x\pm iA_y$
separately,
\begin{align}
  - A_\pm'|^\epsilon_{-\epsilon} &= \frac{4\pi i\omega}{c^2}(\sigma A_\pm \pm i\sigma_1 A_\pm'), \label{bc1a}\\
  A_\pm|^\epsilon_{-\epsilon} &= \frac{4\pi i\omega}{c^2} (\mp i\sigma_1 A_\pm
  + \sigma_2 A'_\pm). \label{bc2a}
\end{align}

We note here some peculiarities of these boundary conditions.  In the
usual problem, only $\sigma$ is nonzero, and the boundary
condition~(\ref{bc1a}) specifies that $A_a$ is continuous across
$z=0$, while its first derivative $A_a'$ has a discontinuity
proportional to the value of $A_a$ at $z=0$.  This leads to a
consistent mathematical problem.  But with nonzero $\sigma_1$ and
$\sigma_2$, Eqs.~(\ref{bc1a}) and (\ref{bc2a}) imply that both $A_a$
and $A'$ are discontinuous, and their discontinuity are proportional
to their values at $z=0$.  The boundary conditions become ambiguous:
at which value of $z$ should the right-hand sides of Eqs.~(\ref{bc1a})
and (\ref{bc2a}) be calculated?

To fully resolve the ambiguity, one needs to solve the scattering
problem for a slab of finite width and carefully take the limit of
small width.  This is done in the Supplementary Materials.  We will
use a shortcut that leads to the correct answer when the jumps of $A$
and $A'$ are small compared to their values.  This requires
\begin{equation}\label{eq:small_sigma1}
  \frac{4\pi}{c^2} \sigma_1\omega \ll 1.
\end{equation}
This condition is usually satisfied.  For example for a sheet a few atomic layer thick, if $\sigma\sim e^2/h$ then $\sigma_1\sim \frac{e^2}h d$.  Then
\begin{equation}
  \frac{4\pi\omega}{c^2}\sigma_1 \sim \frac{e^2}{\hbar c}\frac{d}{\lambda}
  \ll 1.
\end{equation}
When the jump of $A$ is much smaller than $A$, we can replace $A$ on
the right-hand side of the boundary condition by the average values on
the two sides.  This prescription is similar to Griffiths's treatment
of the so-called $\delta'(x)$ potential in quantum
mechanics~\cite{Griffiths1993}.  The boundary conditions can now be
written as
\begin{subequations}\label{bc-symm}
\begin{align}
  - \delta A_\pm' &= \frac{4\pi i\omega}{c^2}(\sigma\bar A_\pm \pm i\sigma_1\bar A_\pm'), \\
  \delta A_\pm  &= \frac{4\pi i\omega}{c^2} (\mp i\sigma_1 \bar A_\pm
  + \sigma_2 \bar A'_\pm).
\end{align}
\end{subequations}
where we have denoted $\bar A\equiv \frac12[A(+\epsilon)+A(-\epsilon)]$.

Let us now consider a problem of scattering of a plane wave onto the
plane.  A plane wave coming from $z=-\infty$ gives rise to a
transmitted and a scattered waves.
\begin{equation}
  A_\pm (z) = \begin{cases}
    e^{ikz} + R_\pm e^{-ikz}, & z<0,\\
    T_\pm e^{ikz}, & z>0.
    \end{cases}
\end{equation}
With the boundary conditions, we find
\begin{subequations}\label{TR}
\begin{align}
  T_\pm &= \frac{1 \pm \tilde\sigma_1 k
    + \frac14 (\tilde\sigma_1^2 - \tilde\sigma\tilde\sigma_2) k^2}
  {1+ \frac12 \tilde\sigma + \frac14
    (2\tilde\sigma_2 - \tilde\sigma_1^2+ \tilde\sigma\tilde\sigma_2) k^2}\,, \\
  R_\pm & =  -\frac12 \frac{\tilde\sigma-\tilde\sigma_2 k^2}
  {1+ \frac12 \tilde\sigma + \frac14
    (2\tilde\sigma_2 - \tilde\sigma_1^2+ \tilde\sigma\tilde\sigma_2) k^2}\,,
\end{align}
\end{subequations}
where
\begin{equation}
  \tilde\sigma = \frac{4\pi}{c} \sigma,
  \quad \tilde \sigma_1 = \frac{4\pi}{c} \sigma_1,
  \quad \tilde \sigma_2 = \frac{4\pi}{c} \sigma_2 .
\end{equation}
As a consistency check of the symmetrized prescription
(\ref{bc-symm}), one can verify that if
$\Re\sigma=\Re\sigma_1=\Re\sigma_2=0$, implying no dissipation, then
$|T|^2 + |R|^2=1$.

Ignoring $\sigma_2$ and assuming $\tilde\sigma_1 k\ll1$, we have
\begin{equation}\label{TRk}
  T_\pm = \frac{{1 \pm \frac{4\pi}c k\sigma_1}}{1+\frac{2\pi\sigma}c}\,,
  \quad R_\pm = - \frac{\frac{2\pi}c\sigma}{1+ \frac{2\pi}c \sigma}\,.
\end{equation}

The Faraday rotation angle is then
\begin{equation}
\label{eq:thetaF}
  \theta_{\rm F} = \frac12\arg \frac{T_+}{T_-} =
  \frac{4\pi\omega}{c^2}  \Im \sigma_1.
\end{equation}
while
the ratio of the absolute value of the transmission amplitude of the
two polarizations (which is different from 1 if there is circular
dichroism) is
\begin{equation}\label{eq:chiral_dichroism}
  \frac{|T_+|^2}{|T_-|^2} = 1 + \frac{16\pi\omega}{c^2} \Re\sigma_1.
\end{equation}
Both Eqs.~(\ref{eq:thetaF}) and (\ref{eq:chiral_dichroism}) do not
require $\sigma$ to be small to be valid, but only
(\ref{eq:small_sigma1}).

In the Supplementary Materials we reproduce the formulas by solving
the Maxwell equations for a slab of finite thickness in the limit of
long wavelength.  Equations~(\ref{eq:thetaF}) and
(\ref{eq:chiral_dichroism}), as written, are correct also in the
presence of a dielectric constant on one or or both sides of the
sheet.

\emph{Metallic state.}---Consider a state with a Fermi surface.  In
the regime
where the incoming photon has energy less than the Fermi energy (typically
that means wavelength in the infrared range)
we can
use the Fermi liquid theory to treat the problem.  In this
description, we have quasiparticles states $|\p,\alpha\>$ where $\p$
is a momentum along the sheet and $\alpha$ is an internal index (for
twisted bilayer graphene $\alpha$ would corresponds to the spin,
valley, and layer degeneracies).  The state has a magnetic moment
$\bm{\mu}^\alpha_\p$, parallel to the sheet.  One can visualize this
magnetic moment as arising from the spiral-like motion of a wave
packet when it moves along the sheet, jumping back and forth between
the two layers.  The energy in a magnetic field $\B$ has the form
(from now on we suppress the $\alpha$ index)
\begin{equation}\label{Ep}
  E_\p = \varepsilon_\p - \bm{\mu}_\p \cdot\B .
\end{equation}
Note that for fields that are constant on the $(x,y)$ plane,
$\B=-\d_z\A\times\z$.  Ignoring Fermi-liquid effects, the kinetic
equation for the distribution function $f_\p$ in the relaxation-time
approximation is
\begin{equation}
  \frac{\d f_\p}{\d t} - e \E \cdot \frac{\d f_\p}{\d \p} =
  - \frac{f_\p - f_0(E_\p)}\tau\,,
\end{equation}
where $f_0(\varepsilon) = [e^{\beta(\varepsilon-\mu)}+1]^{-1}$.
  Linearizing the equation: $f_\p = f_0(\varepsilon_\p) + \delta f_\p$,
we find
\begin{equation}\label{deltaf-sol}
  \delta f_\p = \frac1{1-i\omega\tau}
  \left(-e\tau \E\cdot\v_\p +  \bm{\mu}_\p\cdot\B\right)
  \left(-\frac{\partial f_0}{\partial \varepsilon}\right).
\end{equation}
To compute the current, we use
\begin{equation}
  \mathbf{j} =
  - e\!\int_\p \frac{\d E_\p}{\d \p} (f_0+\delta f_\p),
\end{equation}
where $\int_\p\equiv \int d^2\p/(2\pi)^2$, and the group velocity is
computed using the full dispersion (\ref{Ep}). Inserting
Eq.~(\ref{deltaf-sol}) we find
\begin{equation}
  \mathbf{j} = \frac{e\tau}{1-i\omega \tau}\!\int_\p
  [e(\E\cdot \v_\p) - i\omega (\bm{\mu}_\p\cdot\B)] \v_\p
  \left( -\frac{\d f_0}{\d \epsilon} \right).
\end{equation}
By using the Maxwell equation $\bm{\nabla}\times\E = -c^{-1}\dot\B$, one gets
\begin{equation}
  \sigma = \frac12 \frac{e^2\tau \nu(\varepsilon_{\rm F})}{1-i\omega\tau}
  \< \v_\p^2\>, \quad
  \sigma_1 = -\frac12 \frac{ec\tau\nu(\varepsilon_{\rm F})}{1-i\omega\tau}
  \< \v_\p\cdot \bm{\mu}_\p\>. \label{sigma1-helicity}
\end{equation}
Here $\nu(\varepsilon_{\rm F})$ is the density of state at the Fermi
level and $\<\cdot\>$ means averaging over the Fermi surface:
$\< A \> \equiv \nu^{-1}(\varepsilon_{\rm F})
\int_\p \! A \delta(\varepsilon_\p-\varepsilon_{\rm F})$.

We will call the scalar product of the velocity $\v_\p$ and the
in-plane orbital magnetic moment $\bm{\mu}_\p$ the ``magnetic
helicity'' of the quasiparticle carrying momentum $\p$.  The average
of the magnetic helicity around the Fermi surface
$\<\v_\p\cdot\bm{\mu}_\p\>$ encodes the correlation between between
the direction of motion of the quasiparticles and their magnetic
moment.  Such correlation leads to the appearance of a magnetic moment
density when an electric current flows through the system.
Equation~(\ref{sigma1-helicity}) implies that the twisted conductivity
is related to this locking between velocity and magnetic moment. This locking is also the origin of  the gyrotropic magnetic effect  in 3D \cite{Ma-Pesin2015,Zhong-Moore-Souza2016} and 2D \cite{Wang2020}.

Note that the ratio of $\sigma_1$ and $\sigma$ does not depend on the
mean free time $\tau$,
\begin{equation}
  \frac{\sigma_1}\sigma =
  - \frac ce \frac{\<\v_p\cdot\bm{\mu}_\p\>}{\<\v_\p^2\>} \,,
\end{equation}
and can be read out just from the wave functions of modes at the Fermi
surface.
This has the consequence that if
$\sigma\ll\frac c{2\pi}\approx 137\frac{e^2}h$, circular dichroism, defined
as $\text{CD}=(\mathcal{A}_+-\mathcal{A}_-)/[2(\mathcal{A}_++\mathcal{A}_-)]$ wherer $\mathcal{A}_+$ and $\mathcal{A}_-$ are
the absorption coefficient of the corresponding helicities, is linear
in frequency
\begin{equation}
  \text{CD} = \frac\omega e\frac{\<\v_p\cdot\bm{\mu}_\p\>}{\<\v_\p^2\>} \,.
\end{equation}

In the regime $\omega\tau\ll1$, $\Im \sigma_1\sim\omega$,
so the Faraday rotation angle is of order $\omega^2$.
In the opposite regime $\omega\tau\gg 1$, the conductivities become
purely imaginary and inversely proportional to the frequency $\omega$.  In particular, the
twisted conductivity $\Im\sigma_1\sim \omega^{-1}$.  This means that
the Faraday rotation angle is constant in this regime
\begin{equation}
  \theta_{\rm F} =
  - \frac{2\pi e}c \nu(\varepsilon_{\rm F}) \< \v_\p\cdot\bm{\mu}_\p\>.
\end{equation}

To illustrate the result, consider the case of twisted bilayer
graphene.  Here $\nu(\varepsilon_{\rm F})$ contains a factor of 8 from
the valley, spin, and layer degeneracies.  For large twisting angle
$\theta$ (larger than the angle 1.1$^\circ$ where flat bands appear)
and small Fermi momentum $p_{\rm F}\ll k_\theta=2k_{\rm
  D}\sin\frac\theta2$ where $k_{\rm D}$ is the Dirac momentum, the
quasiparticles are localized on one layer, with small admixture from
the other layer.  The magnetic moment of the quasiparticle is mostly
perpendicular to the momentum, but there is a small component along
the quasiparticle's momentum, leading to a nonzero average magnetic
helicity.  The angle of Faraday rotation in this case is (see the
Supplementary Materials)
\begin{equation}
  \theta_{\rm F} = -24\alpha \frac{v_0}c  \frac{k_{\theta} d}{\hbar}
  \left(\frac{p_{\rm F}}{k_\theta}\right)^2
  \frac{w_{\rm AA}w_{\rm AB}}{(v_0 k_\theta)^2} \,,
\end{equation}
where $\alpha\approx \frac{1}{137}$ is the fine structure constant, $v_0$ is the velocity of the Dirac fermion in single-layer graphene, $w_{\rm AA}$ and $w_{\rm AB}$ are the interlayer coupling
parameters, and $d$ is the distance between the layers.  For example,
at twist angle $\theta=2^\circ$, using the standard numerical values
for the parameters (see, e.g., Ref.~\cite{Tarnopolsky2019})
one gets $\theta_{\rm
  F}\sim 10^{-5}$ when $p_F$ is of the same order of magnitude as
$k_\theta$.  The sign of $\theta_{\rm F}$ is such that the plane of
polarization is rotated in the same direction as the direction, with
respect to which the graphene layer farther from the source is rotated
with respect to the one closer to the source.
Analogously, we obtain $\text{CD}\sim 10^{-4}$ for $\omega\sim v_0k_\theta$.

\emph{Superconducting case.}---We now consider the case of a
superconducting thin layer.  Even when we do not know the microscopic
mechanism for superconductivity in twisted bilayer graphene and other
twisted materials, we can treat the problem of scattering of
long-wavelength electromagnetic waves (with frequency less than the
superconducting gap,
thus typically in the microwave range)
phenomenologically using the Ginzburg-Landau
theory.  Keeping only the phase $\varphi$ of the order parameter, the
Ginzburg-Landau energy functional has the
form~\cite{MineevSamokhin:1994,Edelstein:1996}
\begin{equation}\label{GL}
  H = \!\int\! d^3x\, \delta(z)\biggl[
    \frac{\hbar^2n_s}{2m}(D_a\varphi)^2
    + \kappa n_s B_a D_a\varphi
    \biggr],
\end{equation}
where the covariant derivatives are $D_t\varphi = \d_t \varphi+ \frac{2e}\hbar A_0$, 
$D_a\varphi=\d_a\varphi+\frac{2e}{\hbar c}A_a$, and the index $a$ runs $x,y$.
The term proportional to $\kappa$ in Eq.~(\ref{GL}) is the so-called
``Lifshitz invariant'' term.  One can imagine that this term arises
from a nonminimal coupling of the order parameter $\psi$ with the
electromagnetic field: $\frac i2\kappa
B_a(\psi^\+D_a\psi-D_a\psi^\+\psi)$.  This type of coupling has been
considered previously in the treatment of non-centrosymmetric
superconductors~\cite{MineevSamokhin:1994,Edelstein:1996}.  In that
context, the term is associated with spin-orbit coupling.

For twisted bilayer graphene and other twisted materials, from
symmetry arguments one expects the Lifshitz invariant to appear
independent of the mechanism of superconductivity.  Physically, the
Lifshitz invariant term implies that a moving condensate possesses a
nonzero density of magnetic moment.  This is expected if
the superconductivity is formed by Cooper pairing of quasiparticles with
nonzero average magnetic helicity, which, as we recall, parametrizes
the locking between the velocity and the magnetic moment.

One can couple this action with the electromagnetic field in the bulk
and solve the combined system of equations for the gauge field and the
phase $\varphi$.  Equivalently, one can simply compute the twist
conductivity from Eq.~(\ref{GL}).  Differentiating the action with
respect to $A_a$, and setting all fields to be spatially homogeneous
along the directions of the sheet, in particular $\d_a\varphi=0$
(which can be done when the electromagnetic wave falls perpendicularly
onto the sheet), one obtains a generalized London equation
\begin{multline}
  J^a = -c \frac{\delta H}{\delta A_a} = \delta(z) \left[
    - \frac{(2e)^2n_s}{mc} A_a
    - 2e \frac{\kappa n_s}\hbar B_a \right]\\
  + 2e \frac{\kappa n_s}\hbar \epsilon^{ab} \d_z [\delta(z) A_b],
\end{multline}
from which one finds
\begin{align}
\label{eq:condsup}
  \sigma =  \frac{i(2e)^2 n_s}{m\omega}\,, \quad
  \sigma_1 = \frac{ic}\omega 2e \frac{\kappa n_s}\hbar \,.
\end{align}
Since $\sigma_1\sim\omega^{-1}$, the Faraday rotation angle is
frequency-independent.  This can be seen by applying simple
field-theoretical power counting to the Lifshitz invariant term: as
$B_a$ has dimension 2 and $D_a\varphi$ has dimension one, $\kappa n_s$
must be dimensionless.  Since the Faraday rotation angle is
dimensionless and proportional to $\kappa n_s$, there should be no
frequency dependence.  This is exactly the same behavior as for metals
in the regime $\omega\tau\gg1$.

The uniqueness of the Lifshitz invariant allows one to compute also
the change of the plane of polarization when the incoming light falls
onto the sheet at any angle. This calculation is done in the
Supplementary Materials.

The Lifshitz invariant also leads to another effect---the chiral magnetic Josephson effect in which the Josephson junction is built up from two chiral superconductors linked by a uniaxial ferromagnet. Josephson current appears even with zero phase difference dues to a phase offset \cite{Chernodub:2019,Buzdin:2008}. This phase offset originates from the parity breaking term and is proportional to $\kappa$.  This effect is similar to the chiral magnetic effect in hot QCD \cite{KHARZEEV2004}. 

\emph{Insulating states}.---In the insulating state, there is no
low-energy degree of freedom living on the thin layer, so the
effective action is just a local function of the electromagnetic field
and its derivatives.  Instead of the Lifshitz invariant, the leading
term which breaks the 2D reflection symmetry, but preserve the
combination of 2D reflection and $z\to-z$ symmetry, is
\begin{equation}\label{beta-term}
 \delta S = \int\! dt\,d^3x\,\delta(z)\beta \epsilon^{ab} E_a\d_z E_b,
\end{equation}
where $\beta$ is some constant.  The coefficient $\beta$ has dimension
$-2$, thus one concludes that the Faraday rotation angle behaves like
$\beta\omega^2$ for waves with frequency less than the gap.  This
prediction is independent of the nature of the insulating state,
whether one is dealing with a band insulator or a strongly correlated
insulator.  Calculations similar to the one we have done for the
superconducting case give $\theta_{\rm F}=4\pi\beta(\omega/c)^2$.

To give a physical interpretation to the term~(\ref{beta-term}) in the
effective action, let us imagine immersing a finite piece of thin
twisted insulator into an uniform static parallel electric field
$E_a$.  For simplicity let us also assume that the boundary is also
gapped.  The effective action of this finite piece of material is
described by the same Eq.~(\ref{beta-term}) with $\beta$ replaced by
the function $\beta(x)$, where $\beta(x)=\beta$ inside the piece and 0
outside. The charge density induced by a constant electric field field
is
\begin{equation}
  \rho =
  \frac{\delta S}{\delta A_0} = \delta'(z) \epsilon^{ab} E_a \d_b\beta(x).
\end{equation}
This charge distribution corresponds to zero total charge and dipole
moment, but leads to nonzero value of the electric quadrupole moment
\begin{equation}
  Q^{az} = \!\int\!dz\,d^2x\, z x^a \rho(z,x) = -\beta S\epsilon^{ab} E_b,
\end{equation}
where $S$ is the total area of the piece of material.
(This quadrupole moment is well defined as it is along directions
perpendicular to the direction of the dipole moment.)
Thus $\beta$ is
the coefficient determining the quadrupole polarization induced by an
electric field, which provides an independent way of measuring
$\beta$.

\emph{Conclusion.}---We have constructed the minimal theory describing
the chiral response of a thin sheet of a twisted material, applicable
to twisted bilayer graphene and related materials.  In addition to the
2D electrical conductivity of the sheet, one needs to introduce
another transport coefficient, which we term the ``twisted
conductivity.''  For a metal, the twisted conductivity is proportional
to the average magnetic helicity (i.e., scalar product of the velocity
and the in-plane orbital magnetic moment) of the quasiparticles around
the Fermi line.  For a superconductor, the chiral response is shown to
be related to the Lifshitz invariant, which parametrizes the magnetic
moment density created by a supercurrent.  For the insulating phase,
the chiral response is related to the quadrupole moment induced by an
in-plane electric field.

The discussion in this paper is general and insensitive to the details
of microscopic physics.  It would be interested to see if the chiral
response of the correlated phases is sensitive to the mechanism
underlying these phases.

\emph{Acknowledgments.}---The authors thank Stephen Carr, Van-Nam Do,
Joel Moore, Ashvin Vishwanath, and Grigory Tarnopolsky for discussions and
comments on an earlier version of this manuscript.  DTS is
supported, in part, by the by the U.S. DOE grant
No.\ DE-FG02-13ER41958, a Simons Investigator grant and by the Simons
Collaboration on Ultra-Quantum Matter from the Simons Foundation.  DXN
was supported by Brown Theoretical Physics Center.

\bibliography{twisted}

\onecolumngrid

\newpage
\begin{center}
	\textbf{\large --- Supplementary Material ---\\ Electrodynamics of Thin Sheets of Twisted Material}\\
	\medskip
	\text{Dung Xuan Nguyen and Dam Thanh Son }
\end{center}
\setcounter{equation}{0}
\setcounter{figure}{0}
\setcounter{table}{0}
\setcounter{page}{1}
\makeatletter
\renewcommand{\theequation}{S\arabic{equation}}
\renewcommand{\thefigure}{S\arabic{figure}}
\renewcommand{\bibnumfmt}[1]{[S#1]}

\section{From 3D to 2D}

In this Section, we consider the problem of the scattering of
electromagnetic waves falling perpendicularly onto a slab of a finite
width.  Then taking the limit where the thickness of the slab is much
smaller than the wavelength of the incoming radiation, we reproduce
the formulas for the Faraday rotation angle and circular dichroism
obtained in the main text.

Consider a layer of material, whose response to an external electric
field is given by a nonlocal conductivity $\sigma(\omega; z,z')$
\begin{equation}\label{app:jsE}
  j_a (\omega; z) = \int\!dz'\, \sigma_{ab}(\omega; z,z') E_b(\omega, z').
\end{equation}
We assume that the layer has finite thickness $d$, so $\sigma_{ab}=0$
when $|z|>d/2$ or $|z'|>d/2$, or both.  We will omit the argument
$\omega$ in further formulas.
We want to derive the formula for the transmission and reflection
coefficient in the limit of small $k$, $kd\ll 1$.  

First, we can decompose $\sigma_{ab}$ into symmetric and antisymmetric parts
\begin{equation}
  \sigma_{ab} = \sigma_S \delta_{ab} + \sigma_A \epsilon_{ab}.
\end{equation}
In the helicity basis
\begin{equation}
  j_\pm = j_x \pm ij_y, \qquad E_\pm = E_x \pm iE_y,
\end{equation}
Eq.~(\ref{app:jsE}) can be written as
\begin{equation}
  j_\pm(z) = \int\!dz'\, \sigma_\pm(z,z') E_\pm(z'),
\end{equation}
where we defined
\begin{equation}
  \sigma_\pm = \sigma_S \mp i\sigma_A.
\end{equation}

The symmetry with respect to
reflection around the $y$ axis combined with the exchange of the upper
and lower layers  $z\to-z$ (which is the two-fold rotation
around the $x$ axis) implies
\begin{equation}\label{app:spsm}
  \sigma_+(z,z') = \sigma_-(-z,-z').
\end{equation}
Time reversal invariance implies
$\sigma_{ab}(\omega;z,z')=\sigma_{ba}(\omega;z',z)$, which means
\begin{equation}
  \sigma_+(z,z') = \sigma_-(z',z).
\end{equation}
Combining two symmetries, we find
\begin{equation}\label{app:sigma-symm}
  \sigma_\pm (z, z') = \sigma_\pm (-z', -z).
\end{equation}
From now on we focus on the positive helicity, and drop
the helicity index $+$ in formulas.  Let us also define
\begin{equation}
  \sigma(z) = \int\!dz'\, \sigma(z,z').
\end{equation}
From Eq.~(\ref{app:sigma-symm}) we find
\begin{equation}
  \int\!dz\, \sigma(z,z') = \sigma(-z').
\end{equation}

The Maxwell equation can be written as
\begin{equation}\label{app:Maxwell}
  \d_z^2 A(z) + k^2 A(z) = - i k\! \int\!dz'\, \tilde\sigma(z,z') A(z'),
  \quad \tilde\sigma(z,z') = \frac{4\pi}c \sigma(z,z').
\end{equation}
We will be looking for solution to this equation with the asymptotics
\begin{equation}
  A(z) \to e^{ikz}, \quad z\to +\infty,
\end{equation}
which means that in the opposite limit the asymptotics of $A(z)$ is
\begin{equation}
  A(z) = \frac 1T e^{ikz} + \frac RT e^{-ikz},
\end{equation}
where $T$ and $R$ are the transmission and reflection amplitudes,
respectively. Using Green's function, Eq.~(\ref{app:Maxwell}) can be
written as
\begin{equation}
  A(z) = e^{ikz} - i \!\int\!dz_1\,dz_2\, \theta(z_1-z) \sin k(z_1-z)
  \tilde\sigma(z_1,z_2) A(z_2),
\end{equation}
from which we find
\begin{subequations}\label{app:TR}
\begin{align}
  \frac 1T &= 1 + \frac12 \!\int\!dz_1\, dz_2\, e^{-ikz_1}
  \tilde\sigma(z_1,z_2) A(z_2), \\
  \frac RT &=   - \frac12 \!\int\!dz_1\, dz_2\, e^{ikz_1}
  \tilde\sigma(z_1,z_2) A(z_2).
\end{align}
\end{subequations}

The Maxwell equation~(\ref{app:Maxwell}) can be solved by iteration,
where the solution is presented as an infinite series
\begin{equation}\label{app:iter}
   A(z) = \sum_{n=0}^\infty A^{(n)}(z),
\end{equation}
where we define
\begin{align}
  A^{(0)}(z) &= e^{ikz},\\
  A^{(n)}(z) &= -i \!\int\!dz_1\,dz_2\, \theta(z_1-z) \sin k(z_1-z)
  \tilde\sigma(z_1,z_2) A^{(n-1)}(z_2).
\end{align}
In the limit of small $k$ the expansion~(\ref{app:iter}) is an
expansion over $k$.  First consider $z$ inside the slab, $|z|<d/2$, then
$\sin k(z-z_1)\sim kd$ since $z_1$ has to be inside the slab,
we have
\begin{equation}
  A^{(n)}(z) \sim k A^{(n-1)}(z) \Rightarrow A^{(n)}(z) \sim (kd)^n, \qquad
  |z| < \frac d2 \,.
\end{equation}

Let us first compute $T$ and $R$ to order $k^0$.  We replace in
Eq.~(\ref{app:TR}) $A(z)\to A^{(0)}(z) = e^{ikz}$.  To order $k^0$ we
can replace $e^{\pm ikz_1}$ and $e^{ikz_2}$ by $1$ since $z_1,
z_2\lesssim d$.  We find
\begin{align}
  \frac1T &= 1 + \frac12 \!\int\!dz_1\, dz_2\, \tilde\sigma(z_1,z_2),\\
  \frac RT &= -\frac12 \int\!dz_1\, dz_2\, \tilde\sigma(z_1, z_2).
\end{align}
Comparing with Eqs.~(\ref{TRk}) we can identify
\begin{equation}
  \sigma = \int\!dz_1\, dz_2\, \sigma(z_1,z_2).
\end{equation}

To compute $T$ and $R$ to order $k^1$ we need to replace, in
Eqs.~(\ref{app:TR}) $A(z)$ by $A^{(0)}(z)+A^{(1)}(z)$, where $A^{(0)}$
is computed to next-to-leading order in $k$, and $A^{(1)}$ to leading
order,
\begin{align}
  A^{(0)}(z) &= 1 + ikz + O(k^2), \\
  A^{(1)}(z) &= -ik\! \int\!dz_1\, \theta (z_1-z) (z_1-z)
  \tilde\sigma(z_1) + O(k^2).
\end{align}
We find
\begin{align}
  \frac1T &= 1
  + \frac12\int\!dz\, \tilde\sigma(z)
  - ik \!\int\!dz\, z\tilde\sigma(z)
  - \frac{ik}2\! \int\!dz\,dz'\, \theta(z+z')(z+z')\tilde\sigma(z)\tilde\sigma(z'),
  \\
  \frac RT &=
  - \frac12 \int\!dz\, \tilde\sigma(z)
  + \frac{ik}2\! \int\!dz\,dz'\, \theta(z+z')(z+z')\tilde\sigma(z)\tilde\sigma(z').
\end{align}
Let us now divide $\tilde\sigma(z)$ into symmetric and  antisymmetric parts
\begin{equation}
  \sigma(z) = \sigma_s(z) + \sigma_a(z) ,\qquad
  \sigma_s(z) = \sigma_s(-z), \quad \sigma_a(z) = -\sigma_a(-z)
\end{equation}
Then we find
\begin{align}
\label{eq:1/T}
  \frac1T &= \left[ 1 + \frac12\int\!dz\, \tilde\sigma_s(z)\right]
  \left[ 1 - ik\!\int\!dz\, z\tilde\sigma_a(z) \right]
  -\frac{ik}2\! \int\!dz\,dz'\, \theta(z+z')(z+z')
  [\tilde\sigma_s(z)\tilde\sigma_s(z')+ \tilde\sigma_a(z)\tilde\sigma_a(z')],\\
  \label{eq:R/T}
  \frac RT &= - \frac12 \int\!dz\, \tilde\sigma(z)
  \left[ 1 - ik\!\int\!dz\, z\tilde\sigma_a(z) \right]
  +\frac{ik}2\! \int\!dz\,dz'\, \theta(z+z')(z+z')
  [\tilde\sigma_s(z)\tilde\sigma_s(z')+ \tilde\sigma_a(z)\tilde\sigma_a(z')].
\end{align}
If one now identifies
\begin{align}
  \tilde\sigma &= \int\!dz\, \tilde\sigma(z) - ik\! \int\!dz\,dz'\, \theta(z+z')(z+z')
  [\tilde\sigma_s(z)\tilde\sigma_s(z')+ \tilde\sigma_a(z)\tilde\sigma_a(z')] ,\\
  \tilde\sigma_1 &= i\!\int\!dz\, z\tilde\sigma_a(z),
\end{align}
then equations \eqref{eq:1/T} and \eqref{eq:R/T} can be written as
\begin{align}
  \frac1T &= \left( 1+\frac{\tilde\sigma}2 \right)(1-k\tilde\sigma_1),\\
  \frac RT &= -\frac{\tilde\sigma}2 (1-k\tilde\sigma_1),
\end{align}
which coincide, to order $O(k)$, with Eq.~(\ref{TRk}).
For negative helicity $-$, it is easy to find from
Eq.~(\ref{app:spsm}) that $\tilde\sigma$ remains the same, but
$\tilde\sigma_1$ flips sign.

\section{Twisted bilayer graphene}

We compute the orbital magnetic moment of the quasiparticles twisted
bilayer graphene.  We use the continuum theory \cite{Bistritzer2011, Tarnopolsky2019}.  We follow the notation of Ref.~\cite{Tarnopolsky2019}. The system consists of two layers; the upper lay is rotated clockwise by an angle $\theta>0$ with respect to the lower layer. The single-particle Hamiltonian is
\begin{equation}
  H = \begin{pmatrix}
    -i v_0 \bm{\sigma}_{\theta/2}\cdot \bm{\nabla} & T(\x)\\
    T^\+(\x) & -iv_0 \bm{\sigma}_{-\theta/2}\cdot\bm{\nabla} 
    \end{pmatrix} .
\end{equation}
Here we denoted
\begin{equation}
  \bm{\sigma}_{\theta/2} =
  e^{-i\theta\sigma_z/4} \begin{pmatrix} \sigma_x \\ \sigma_y \end{pmatrix}
  e^{i\theta\sigma_z/4}.
\end{equation}
We also used
\begin{equation}
  T(\x) = \sum_{a=1}^3 T_a e^{-i\q_a \cdot \x},
\end{equation}
where we defined the momentum $\mathbf{q}_a$ as 
\begin{equation}
  \q_1 = \begin{pmatrix} 0 \\ -1 \end{pmatrix} k_\theta, \qquad
  \q_2 = \begin{pmatrix} {\sqrt3}/2 \\ 1/2 \end{pmatrix} k_\theta, \qquad
  \q_3 = \begin{pmatrix} -{\sqrt3}/2 \\ 1/2 \end{pmatrix} k_\theta, \qquad
\end{equation}
and
\begin{equation}
  T_1 =  \begin{pmatrix} \waa & \wab \\ \wab & \waa \end{pmatrix}, \quad
  T_2 =  \begin{pmatrix} \waa & \wab e^{-2\pi i/3} \\ \wab e^{2\pi i/3}  & \waa \end{pmatrix}, \quad
  T_3 =  T_2^\+.
\end{equation}
The matrix $T_a$ can be written as
\begin{equation}
  T_a = \waa + \wab (\z \times \hat\q_a) \cdot \bm{\sigma},
\end{equation}
where $\hat \q_a = \q_a/k_\theta$ is the unit vector directed along
the direction of $\q_a$. It is convenient to rotate the basis to transform the Hamiltonian into
\begin{equation}
  H \to \begin{pmatrix} e^{i\frac\theta4\sigma_z} & 0\\
    0 & e^{-i\frac\theta4\sigma_z} \end{pmatrix}
  H
  \begin{pmatrix} e^{-i\frac\theta4\sigma_z} & 0\\
    0 & e^{i\frac\theta4\sigma_z} \end{pmatrix}
  = \begin{pmatrix}
    -i v_0 \bm{\sigma}\cdot \bm{\nabla} & T_\theta(\x)\\
    T_\theta^\+(\x) & -iv_0 \bm{\sigma}\cdot\bm{\nabla} 
  \end{pmatrix},
\end{equation}
where
\begin{equation}
  T_\theta = \sum_a T_\theta^a e^{-i\q_a\cdot \x},\quad
  T_\theta^a = \waa e^{i\theta\sigma_z/2}  +
  \wab (\z\times \hat \q_a) \cdot \bm{\sigma} .
\end{equation}

To find the orbital magnetic moment of a state $|\psi\>$, one can turn
on a small in-plane magnetic field $\B$ by turning on opposite gauge
potentials on the two layers,
\begin{equation}
  \A_u = - \frac d2 (\z \times \B), \quad
  \A_d =  \frac d2 (\z \times \B),
\end{equation}
where $u$ and $d$ indicates upper and lower layers, and $d$ is the
distance between the two layers.  The interaction of the system with
the gauge field $\A_u$ and $\A_d$ is given by
\begin{equation}
  \delta H = - \frac1c (\J_u\cdot \A_u + \J_d\cdot \A_d)
  = \frac1{2c} (\J_u - \J_d) \cdot (\z\times\B).
\end{equation}
By comparing this with $\delta H=-\bm{\mu}\cdot\B$, we find
\begin{equation}
  \bm{\mu} = \frac d{2c}\z \times (\J_u-\J_d)
  = -\frac {ed}{2c} \z \times (\bm{\sigma}_u - \bm{\sigma}_d). 
\end{equation}
In the regime $\waa,\wab\ll v_0 k_\theta$, we can use perturbation
theory over $T$.  To first order in perturbation the eigenstate of a Hamiltonian $H=H_0+H_1$ is
\begin{equation}
  |0\>' = |0\> - \sum_{n\neq0} \frac{|n\>\<n|H_1|0\>}{E_n-E_0} + \mathcal{O}(T^2).
\end{equation}
If $|0\>$ is a state on the upper layer with $p\approx 0$, then $|n\>$
are states on the lower layer with momentum $\p+\q_a\approx q_a$,
$a=1,2,3$.  Thus we need to consider only 8 bands $|0,\alpha\>$ and
$|a,\alpha\>$ with $\alpha=1,2$ being the spinor index.  The
unperturbed Hamiltonian is
\begin{equation}
  \<0,\alpha| H_0 |0,\beta\> = v_0 \p\cdot\bm{\sigma}_{\alpha\beta}, \quad
  \<a,\alpha| H_0 |b,\beta\> = v_0 (\q_a+\p)\cdot\bm{\sigma}_{\alpha_\beta} \delta_{ab} ,
\end{equation}
while the perturbation part of the Hamiltonian is
\begin{equation}
   \<0,\alpha |H_1 | a,\beta\> = (T^\theta_a)_{\alpha\beta} .
\end{equation}
The expectation value of the magnetic moment is then
\begin{equation}
   \bm{\mu} = -\frac d{2c} \z\times \left( \<0|\bm{\sigma}_u|0\>
  - \sum_{mn} \frac{ \<0| H_1 |m\> \< m| \bm{\sigma}_d |n\>\< n| H_1 |0\>}
    {(E_m-E_0)(E_n-E_0)} \right).
\end{equation}
Note that $\< 0 |\bm{\sigma} |0\> = \hat\p$, and therefore does not
contribute to the magnetic helicity.  Thus, we only need to compute
the second term inside the bracket
In each sector with fix $a$, $H_0$ has two energy eigenvectors
\begin{equation}
  H_0 |a, \pm\> = \pm v_0 p_a |a,\pm\>, \qquad p_a \equiv |\q_a+\p|.
\end{equation}
The sum over intermediate states in $a$-sector gives
\begin{equation}
  S_a \equiv \sum_{\alpha=\pm} \frac{|a,\alpha\>\<a,\alpha|}{E_\alpha-E_0}
  = \frac1{2v_0(p_a-p)}
  ( 1 + \bm{\sigma}\cdot\hat\p_a)
  -  \frac1{2 v_0(p_a+p)}
  ( 1 - \bm{\sigma}\cdot\hat\p_a).
\end{equation}
Putting everything into the expression for the orbital magnetic moment,
we then find
\begin{equation}\label{eq:vmuT}
  \v_\p\cdot\bm{\mu}_\p =
  \frac d{2c}\v_\p\cdot\sum_{a=1}^3 \<0 | T_a^\theta S_a (\z\times\bm{\sigma}) S_a
     (T_a^\theta)^\dagger | 0\> .
\end{equation}
For $p\ll k_\theta$, this can be evaluated to give
\begin{equation}
  \< \v_\p\cdot\bm{\mu}_\p\> = \frac{6 w_{\rm AA} w_{\rm AB}}{(v_0k_\theta)^2}
  \frac p{k_\theta}\frac{ed v_0^2}{2c} + \mathcal{O}(p^2).
\end{equation}
The presence of a slip (vector $\mathbf{d}$ in
Ref.~\cite{Bistritzer2011}) changes the overall phase of the matrix
$T(\x)$ and hence according to Eq.~(\ref{eq:vmuT}) does not change the
result for the magnetic helicity.

\section{Faraday rotation of the superconducting phase with a finite incident angle $\alpha$}
\label{sec:FRtheta}
\subsection{Setup and equation of motions}
\label{sec:setup}
\begin{figure}
	\begin{center}
		\includegraphics[width=0.4\textwidth]{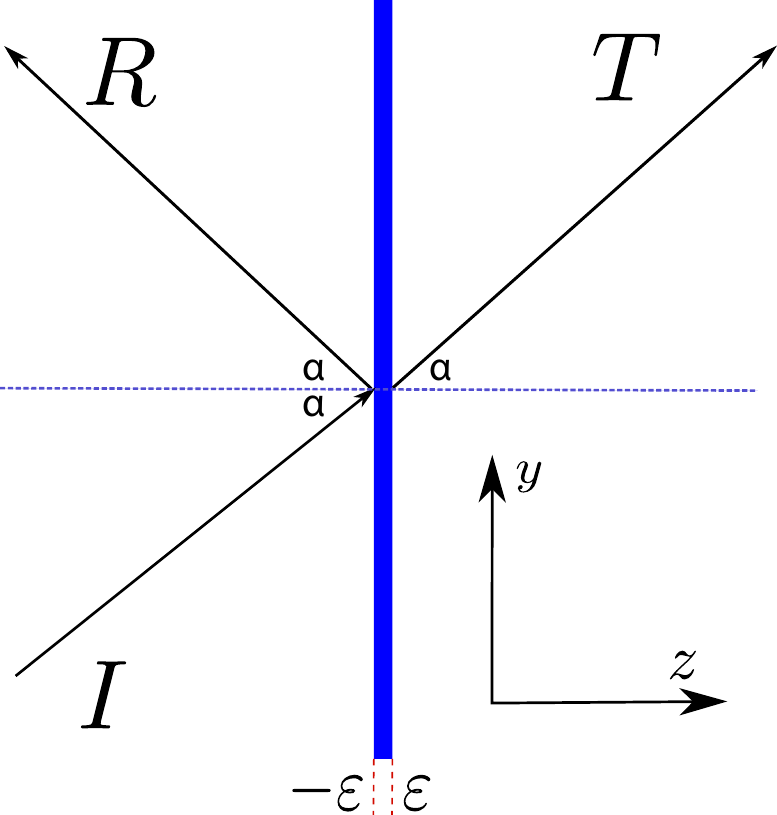}
		\caption{Light scattering diagram}
		\label{fig:Diag}
	\end{center}  
\end{figure}

We consider a thin chiral superconductor with thickness $2\ve$. The scattering light lives on the $(y,z)$ plane and makes an angle $\alpha$ with the $z$ axis as shown in Fig \ref{fig:Diag}. We recall the action of a thin chiral superconductor interacting with the electromagnetic field 
\begin{align}
\label{eq:s2}
\cS=\int\! d^4x \left\{-\frac{1}{16 \pi}F_{\mu\nu}F^{\mu\nu}
+\delta(z)\left[ \frac{f^2}2 (D_t\varphi)^2 - \frac{ n_s}{2m} (D_a\varphi)^2 -n_s \kappa B_a D_a \vp \right]\right\},
\end{align}
where we used $F_{\mu\nu}=\partial_\mu A_\nu-\partial_\nu A_\mu$ and $\partial_0=\frac{1}{c}\partial_t$. We chose subscripts $a,b$ to denote inplane spacial directions.  We chose the Coulomb gauge $A_0=0$, instead of $\partial_a \vp=0$ for $\alpha=0$ as in the main text, the equation of motion of $\varphi$ is non-trivial. The components of incident, reflection and transmission lights has the following ansatz
\begin{align}
&A_x^I e^{-i (\omega t -\mathbf{k}\cdot \mathbf{r})}, \quad A_y^I  e^{-i (\omega t -\mathbf{k}\cdot \mathbf{r})},\quad A_z^I  e^{-i (\omega t -\mathbf{k}\cdot \mathbf{r})},\\
&A_x^R e^{-i (\omega t -\mathbf{k}'\cdot \mathbf{r})}, \quad A_y^R  e^{-i (\omega t -\mathbf{k}'\cdot \mathbf{r})}, \quad A_z^R  e^{-i (\omega t -\mathbf{k}'\cdot \mathbf{r})},\\
&A_x^T e^{-i (\omega t -\mathbf{k}\cdot \mathbf{r})}, \quad A_y^R  e^{-i (\omega t -\mathbf{k}\cdot \mathbf{r})}, \quad A_z^R  e^{-i (\omega t -\mathbf{k}\cdot \mathbf{r})}.
\end{align}
where we denote $\mathbf{k}=(0,k_y,k_z)$ and $\mathbf{k}'=(0,k_y,-k_z)$, with $k_y=k \sin \alpha, k_z=  k \cos \alpha$. Our task is finding $A^R_i$ and $A^T_i$ in term of $A^I_i$. The equation of motion (e.o.m) of $A_0$ gives
\begin{equation}
\label{eq:A_0}
0=\frac{1}{c}\partial_t \partial_i A_i +2 \delta(z) f^2 \partial_t \varphi.
\end{equation}
Equation \eqref{eq:A_0} is nothing but the Gauss law which implies $\nabla_i A_i=0$ outside the superconductor. 
The e.o.m of $\vp$ is
\begin{equation}
\label{eq:vp}	
0=\delta(z)\left[-f^2 \d_t^2 \vp  +\frac{n_s}{m}\d_a D_a \vp +n_s \kappa \d_a B_a\right].
\end{equation}
Finally the e.o.m of $A_i$ are
\begin{equation}
\label{eq:Az}
0=-\frac{1}{4 \pi c^2}\d_t^2 A_z -\frac{\epsilon^{ab}}{4\pi}\d_a B_b - n_s \kappa \delta(z) \epsilon^{ab}\d_a D_b \vp ,
\end{equation}
\begin{equation}
\label{eq:Ax}
0=-\frac{1}{4 \pi c^2}\d_t^2 A_x-\frac{(\d_y B_z-\d_z B_y)}{4\pi}-\delta(z)\left(\frac{2 n_s e}{m \hbar c}D_x \vp+2\frac{n_s e}{ \hbar c} \kappa B_x\right)+n_s \kappa \d_z \left(\delta(z) D_y \vp\right),
\end{equation}
\begin{equation}
\label{eq:Ay}
0=-\frac{1}{4\pi c^2}\d_t^2 A_y-\frac{(\d_z B_x-\d_x B_z)}{4\pi}-\delta(z)\left(\frac{2 n_s e}{m \hbar  c}D_y \vp+2\frac{n_s e}{\hbar  c} \kappa B_y\right)-n_s \kappa \d_z \left(\delta(z) D_x \vp\right).
\end{equation}
The combination of Gauss law \eqref{eq:A_0} and the e.o.m for $A_i$  gives us
\begin{equation}
\label{eq:Maxwell}
\frac{1}{c^2}\partial_t^2 A_i-\Laplace A_i=0
\end{equation}
outside the superconductor. It is satisfied automatically with photon's dispersion relation $\omega=c|\mathbf{k}|$. In current setup, one can replace $\d_x \rightarrow 0$. We also replace $\d_y \rightarrow  i k_y$ and $\d_t \rightarrow -i\omega$ to convert above e.o.m to the momentum space.  We employ the long-wave length limit approximations
\begin{equation}
A_i(0)=\frac{1}{2}(A_i(\ve)+A_i(-\ve)), \qquad A'_i(0)=\frac{1}{2}(A'_i(\ve)+A'_i(-\ve)),
\end{equation}
with the notation $A'_i=\partial_z A_i$. Integrating equations \eqref{eq:A_0}--\eqref{eq:Ay} from $-\varepsilon$ to $\ve$, we obtain 
\begin{equation}
\label{eq:A01}
-i\omega \left( (A_z (\ve)-A_z(-\ve))+2c 4\pi f^2 \vp \right)=0,
\end{equation}
\begin{equation}
\label{eq:vp1}
f^2 \omega^2 \vp-\frac{n_s}{m}\left[k_y^2 \vp -\frac{i}{c} k_y  \left(A_y(\ve)+A_y(-\ve)\right)\right]+i k_y n_s  \kappa\frac{1}{2}\left(A'_x(\ve)+A'_x(-\ve)\right)=0,
\end{equation}
\begin{equation}
\label{eq:Az1}
-i k_y (A_y(\ve)-A_y(-\ve))+i k_y \frac{4\pi n_s e}{\hbar c} \kappa (A_x(\ve)+A_x(-\ve)) =0,
\end{equation} 
\begin{equation}
\label{eq:Ax1}
A'_x(\ve)-A'_x(-\ve)-4\pi\left\{\frac{2n_s e^2}{m \hbar^2 c^2}(A_x(\ve)+A_x(-\ve))-\frac{n_s e}{\hbar c} \kappa \left[-i k_y (A_z(\ve)+A_z(-\ve))+(A'_y(\ve)+A'_y(-\ve))\right]\right\}=0,
\end{equation}    
\begin{equation}
\label{eq:Ay1}
A'_y(\ve)-A'_y(-\ve)-i k_y\left(A_z(\ve)-A_z(-\ve)\right)-4\pi\left\{\frac{2n_s e}{m \hbar c}\left[ik_y \vp+\frac{e}{\hbar c}(A_y(\ve)+A_y(-\ve))\right]+\frac{n_s e}{\hbar c} \kappa (A'_x(\ve)+A'_x(-\ve))\right\}=0.
\end{equation}    
Multiplying equations \eqref{eq:Ax}-\eqref{eq:Ay} by $z$ then integrating from $-\ve$ to $\ve$, we obtain 
\begin{equation}
\label{eq:Ax2}
-(A_x(\ve)-A_x(-\ve))+4\pi n_s \kappa \left[-i k_y \vp - \frac{e}{\hbar c}(A_y(\ve)+A_y(-\ve))\right]=0,
\end{equation}

\begin{equation}
\label{eq:Ay2}
-(A_y(\ve)-A_y(-\ve))+4\pi\frac{n_s e}{\hbar c} \kappa  (A_x(\ve)+A_x(-\ve))=0.
\end{equation}
We now have 9 equations including  \eqref{eq:A01}-\eqref{eq:Ay2},and the Gauss law for $z<-\ve$ and $z>\ve$
\begin{equation}
A_y^R k_y-A_z^R k_z=0, \quad A_y^T k_y+A_z^T k_z=0,
\end{equation}
to find 7 unknowns $\vp, A^R_i, A^T_i$ from $A^I_i$ and $k_y,k_z$ with the condition $A_y^I k_y+A_z^I k_z=0$ implied by the Gauss law. One can check that 2 of above  9 equations are redundant. We can solve uniquely $A_i^R, A_i^T$ from $A_i^I$. We will summarize the results in the next subsection.
\subsection{Transmission of linear polarized light}
Due to the Gauss law, we define in-plane vector potential as 
\begin{equation}
A_{in}^I=\frac{A_y^I}{\cos \alpha}=-\frac{A_z^I}{\sin \alpha} ,\quad
A_{in}^R=\frac{A_y^R}{\cos \alpha}=\frac{A_z^R}{\sin \alpha} ,\quad
A_{in}^T=\frac{A_y^T}{\cos \alpha}=-\frac{A_z^T}{\sin \alpha}.
\end{equation}
We also define the out of plane vector potential as
\begin{equation}
A_{out}^I=-A_x^I,\quad
A_{out}^R=A_x^R,\quad
A_{out}^T=-A_x^T.
\end{equation}
The polarization function of the transmission light is
\begin{equation}
\label{eq:pol}
\mathcal{P}^{T}=\frac{A_{out}^{T}}{A_{in}^T}.
\end{equation}
From the definition \eqref{eq:pol}, we see that real (pure imaginary) $\mathcal{P}$ correspond to linear (elliptical) polarized light. Especially, $\mathcal{P}=i$ ($\mathcal{P}=-i$) corresponds to   \textbf{right-hand circle} (\textbf{left-hand circle}) polarized light. We consider the linear polarized incident light with either in-plane polarization or out-of-plane. We calculate the polarization function using equations in the previous section and quote the result at the \textbf{leading orders} in the momentum $k$: 

\textbf{In-plane incident light   ($A_{in}^I \neq 0, A_{out}^I=0$)}\\

We obtain the result of polarization functions as 
\begin{equation}
\label{eq:PTi}
\mathcal{P}^{T}=\frac{\kappa  m c\left(-64\pi^2 \frac{f^2}{c^2} \frac{n_s e^2}{m \hbar ^2 c^2} \cos (\alpha )+i k \sin ^2(\alpha ) \left( \frac{4\pi f^2}{c^2} \left(16\pi^2\kappa ^2  \frac{n_s^2 e^2}{\hbar^2 c^2}+1\right)-\frac{4\pi n_s e^2}{m \hbar^2 c^2}\right)\right)}{2  \frac{f^2}{c^2}  \left(16\pi^2\kappa ^2 \frac{n_s^2 e^2}{\hbar^2 c^2}-1\right)+2\frac{n_s e^2}{m \hbar^2 c^2} \sin^2
	(\alpha )}+ \mathcal{O}(k^2).
\end{equation}

\textbf{Out-of-plane incident light ($A_{in}^I = 0, A_{out}^I \neq 0$)}\\

We obtain the result of polarization functions as 
\begin{equation}
\label{eq:PTo}
1/\mathcal{P}^{T}=\frac{\frac{\kappa  m \hbar c}{e\cos^2(\alpha)} \left(64 \pi^2  \frac{f^2}{c^2} \frac{n_s e^2 }{m \hbar^2 c^2}\cos (\alpha)+i k \sin^2 (\alpha )  \left( 4\pi\frac{f^2}{c^2} \left(16\pi^2\kappa ^2  \frac{n_s^2 e^2}{\hbar^2 c^2}+1\right)-\frac{4\pi n_s e^2}{m \hbar^2 c^2}\right)\right)}{2  \frac{f^2}{c^2} \left(16\pi^2 \kappa ^2
	\frac{n_s^2 e^2}{\hbar^2 c^2}-1\right)} \,.
\end{equation}

From equation \eqref{eq:PTi} and \eqref{eq:PTo}, at the incident angle $\alpha=0$, we obtain the Faraday rotation angle of the linear polarized light (both in-plane and out-of-plane)
\begin{equation}
\label{eq:FR1}
\theta_{F}=\arctan\left(\frac{8\pi  \kappa \frac{n_s e}{\hbar c}}{1-16\pi^2\kappa^2 \frac{n_s^2 e^2}{\hbar^2 c^2}}\right)\approx 8\pi \frac{\kappa n_s e}{\hbar c},
\end{equation}
which is the same as the Faraday rotation angle in the main text for the superconducting phase by combining Eqs. \eqref{eq:condsup} and \eqref{eq:thetaF} . With a general value of $\alpha$, the transmission light is neither linearly nor elliptically polarized.

\end{document}